\documentclass[journal,onecolumn]{IEEEtran}
\usepackage{amsmath}
\usepackage{booktabs}
\usepackage{latexsym}
\usepackage{mathrsfs}
\usepackage{amsthm}
\usepackage{amssymb}
\usepackage{amsfonts}
\usepackage{amsbsy}

\usepackage[shortlabels]{enumitem}
\usepackage{url}
\usepackage{array}
\usepackage{pdflscape}
\usepackage{xcolor}
\usepackage{stmaryrd}
\usepackage{verbatim}
\usepackage{epsfig}
\usepackage{caption}
\usepackage{pmat}
\newcommand{\Ff}{{\mathbb F}}

\newcommand\rank{\operatorname{rank}}   %rank

\newcommand\cc{{\mathcal C}}        %

\newcommand\cF{{\mathbf c}}

\theoremstyle{plain}% default
\newtheorem{thm}{Theorem}%[section]
\newtheorem{lem}[thm]{Lemma}

\newtheorem{cor}{Corollary}
\theoremstyle{definition}

\newtheorem{example}{Example}
\theoremstyle{remark}
\newtheorem{remark}{Remark}
\usepackage[colorlinks=true]{hyperref}

\begin{document}

\title{Three New Infinite Families of Optimal Locally Repairable Codes from Matrix-Product Codes}

\author{\centerline{Gaojun Luo, Martianus Frederic Ezerman, and San Ling}
\thanks{G. Luo, M. F. Ezerman, and S. Ling are with the School of Physical and Mathematical Sciences, Nanyang Technological University, 21 Nanyang Link, Singapore 637371, e-mails: $\{\rm gaojun.luo, fredezerman, lingsan\}$@ntu.edu.sg.}
\thanks{The authors are supported by Nanyang Technological University Research Grant No. 04INS000047C230GRT01.}
\thanks{This work has been submitted to the IEEE for possible publication. Copyright may be transferred without notice, after which this version may no longer be accessible.}
%\thanks{Copyright (c) 2021 IEEE. Personal use of this material is permitted. However, permission to use this material for any other purposes must be obtained from the IEEE by sending a request to pubs-permissions@ieee.org.}
}

%\markboth{}%
%{Shell \MakeLowercase{\textit{et al.}}: Bare Demo of IEEEtran.cls for IEEE Transactions on Magnetics Journals}

\maketitle

\begin{abstract}
Locally repairable codes have become a key instrument in large-scale distributed storage systems. This paper focuses on the construction of locally repairable codes with $(r,\delta)$-locality that achieve the equality in the Singleton-type bound. We use matrix-product codes to propose two infinite families of $q$-ary optimal $(r,\delta)$ locally repairable codes of lengths up to $q^2+q$. The ingredients in the matrix-product codes are either linear maximum distance separable codes or optimal locally repairable codes of small lengths. Further analysis and refinement yield a construction of another infinite family of optimal $(r,\delta)$ locally repairable codes. The codes in this third family have unbounded lengths not divisible by $(r+\delta-1)$.

The three families of optimal $(r,\delta)$ locally repairable codes constructed here are new. Previously constructed codes in the literature have not covered the same sets of parameters. Our construction proposals are flexible since one can easily vary $r$ and $\delta$ to come up with particular parameters that can suit numerous scenarios.
\end{abstract}


\section{Introduction}\label{sec:intro}

The importance of large-scale distributed storage systems in our era of exponentially growing data is evident. Examples of their deployment by large corporations come in names such as Google's {\tt Bigtable}, Microsoft Azure' {\tt Blob Storage}, and Amazon's {\tt S3}. Such systems are ubiquitous to ensure data availability in the event that some of the nodes have gone down, for whatever reason. Erasure codes were introduced to retrieve failed nodes and attain better repair efficiency. Among erasure codes, maximum distance separable (MDS) codes are appealing. They can repair the maximum number of node failures for a given redundancy. For a small amount of node failures, however, MDS coding schemes are low in efficiency. To repair failed nodes more efficiently, locally repairable codes (LRCs) was proposed in \cite{LRC}.

\subsection{Codes with locality}

Let $q$ be a prime power and let $\Ff_q$ be the finite field with $q$ elements. If a linear code $\cc$ over $\Ff_q$ is a $k$-dimensional subspace of $\Ff_q^n$ and has the minimum Hamming distance $d$, then we denote its parameters by $[n,k,d]_q$. Such a code $\cc$ is \emph{maximum distance separable} (MDS) if $d = n-k+1$. Following Gopalan {\it et al.} in \cite{LRC}, we say that the $i^{\rm th}$ symbol $c_i$, for $1\leq i\leq n$, of $\cc$ has \emph{locality} $r$, with $1\leq r\leq k$, if it can be recovered from at most $r$ other code symbols in $\cc$. The code $\cc$ is an $r$-LRC with parameters $[n,k,d]_q$ if all of its symbols have locality $r$.

To address the problem of multiple device failures in practical scenarios, Prakash {\it et al.} in \cite{MLRC} extended the concept of $r$-locality to $(r,\delta)$-locality. In their setup, every symbol can be \emph{recovered locally} in the presence of \emph{additional} $\delta-2$ erasures. For an $[n,k,d]_q$ code $\cc$, the $i^{\rm th}$ symbol $c_i$, again for $1\leq i\leq n$, of $\cc$ is said to have \emph{$(r,\delta)$-locality} if there exists a subset $S_i \subseteq [n] := \{1,2,\ldots,n\}$ containing $i$ and a punctured code $\cc|_{S_i}$ such that the length $|S_i| \leq r + \delta-1$ and the distance ${\rm d}(\cc|_{S_i}) \geq \delta$, where $\cc|_{S_i}$ is the code $\cc$ punctured on the coordinate set $[n]\setminus S_i$ by deleting the components indexed by the set $[n] \setminus S_i$ in each codeword of $\cc$. The code $\cc$ is called an $(r,\delta)$-LRC if all symbols possess $(r,\delta)$-locality.

It is clear that, if $\delta=2$, then the $(r,\delta)$-locality in \cite{MLRC} reduces to the $r$-locality in \cite{LRC}. Let $\lceil \cdot \rceil$ be the ceiling function. In \cite{MLRC}, the trade-off, called the Singleton-type bound, among the parameters $n$, $k$, $d$, $r$, and $\delta$ of an $(r,\delta)$-LRC is given as
\begin{equation}\label{LRCbound}
d\leq n-k+1-\left(\left\lceil\frac{k}{r}\right\rceil-1\right)(\delta-1).
\end{equation}
An $(r,\delta)$-LRC achieving this bound with equality is said to be \emph{optimal}. Finding explicit constructions of optimal $(r,\delta)$-LRCs is both of theoretical and practical significance. This general problem has attracted considerable attention, as exemplified by the respective works in \cite{LRCC2, QiuLRC, LRCXing, LRCJin, LRCZhang1, LRCZhang2, Cai, ChenBC, MLRC, LRC, LRC1, LRC2, LRC3, LRC4, LRC5, LRC6, LRC7, LRC8, LRC9, LRC10, LRC11}.

\subsection{Known results on optimal $(r,\delta)$-LRCs}\label{subsec:known}

An early construction of optimal LRCs, known as {\it pyramid codes}, was introduced in \cite{LRC10}. Based on Gabidulin codes, Silberstein {\it et al.} in \cite{LRC11} presented another class of optimal LRCs. In \cite{LRC1}, a construction of $q$-ary optimal LRCs of length $n \leq q$ was derived from Reed-Solomon codes. This construction was extended in \cite{LRC3} by using the automorphism group of rational function fields. Using elliptic curves, Li, Ma, and Xing in \cite{LRC4} constructed some $q$-ary optimal LRCs of lengths up to $q + \sqrt{q}$. For practical applications, it is desirable to construct optimal LRCs of a large length relative to the field size $q$. Like in the case of MDS codes, determining the maximum length of optimal LRCs is an interesting question. Guruswami, Xing, and Yuan in \cite{LRClength} provided an upper bound on the lengths of optimal LRCs. They also proposed several $q$-ary optimal LRCs of lengths superlinear in $q$. Taken collectively, for minimum distances $d \leq 10$, optimal LRCs of superlinear lengths have been constructed in many works, {\it e.g.}, those in \cite{LRC5, LRClength, LRCJin, LRCXing, LRCShang}.

Cyclic and constacyclic codes are powerful ingredients in the construction of optimal $(r,\delta)$-LRCs due to their nice algebraic structure. Some families of cyclic or constacyclic optimal $(r,\delta)$-LRCs over $\Ff_q$ of length $n \mid (q+1)$ or $n \mid (q-1)$ were built in \cite{LRC2,LRCC2,QiuLRC}. Optimal $(r,\delta)$-LRCs produced by cyclic codes, in particular, have flexible parameters $r$ and $\delta$. However, it is difficult to construct cyclic optimal $(r,\delta)$-LRCs of superlinear lengths. Prior to our present work, only a few families of optimal cyclic $(r,\delta)$-LRCs were known. They have unbounded lengths but small minimum distances \cite{Fang,LRC5}. In \cite{Cai}, the maximum length of optimal $(r,\delta)$-LRCs was determined when $(r+\delta-1)$ is a divisor of the length $n$. Employing combinatorial structures, some families of optimal $(r,\delta)$-LRCs with superlinear lengths were proposed very recently in \cite{Cai,LRCKong}. Another approach to designing optimal $(r,\delta)$-LRCs of superlinear length is by a direct construction of suitable parity-check matrices, as was done, {\it e.g.,} in \cite{ChenBC,LRCZhang1,LRCZhang2}.

\subsection{Our contributions and techniques}

This paper aims at constructing infinite families of optimal $(r,\delta)$-LRCs over $\Ff_q$ of superlinear lengths. The techniques and results can be summarized as follows.
\begin{itemize}
\item[1.]  Matrix-product codes (see Section \ref{subsec:mpcodes}) were introduced in \cite{MPC1,MPC2} to construct longer codes from known shorter ones. It was demonstrated in \cite{MPC3,MPC4,QC} that some quasi-cyclic codes as well as repeated-root cyclic and constacyclic codes can be expressed as matrix-product codes. In this paper, we use matrix-product codes to construct optimal $(r,\delta)$-LRCs of superlinear lengths. This approach, to the best of our knowledge, has never been attempted before.

Starting from nested linear codes $\cc_M\subseteq \cdots \subseteq\cc_1$, we show that the linear code $\cc_1$ and the matrix-product code built from $\cc_1,\cdots,\cc_M$ possess the same $(r,\delta)$-locality. In other words, if $\cc_1$ is an $(r,\delta)$-LRC, then so is the matrix-product code. By selecting MDS codes or optimal $(r,\delta)$-LRCs as the nested codes $\cc_1,\ldots,\cc_M$, we provide an easily computable criterion for the corresponding matrix-product code to be optimal with respect to the bound in (\ref{LRCbound}). Using the criterion, we obtain a huge number of $q$-ary optimal $(r,\delta)$-LRCs for any fixed $q$. These codes are presented in Table \ref{table3}. In particular, the total number of $q$-ary optimal $(r,\delta)$-LRCs obtained based on the criterion increases nonlinearly as $q$ grows larger.

\item[2.] We construct three infinite families of optimal $(r,\delta)$-LRCs with superlinear lengths.

The codes in the first two families of optimal $q$-ary $(r,\delta)$-LRCs (Cor. \ref{cor1} and Cor. \ref{cor2}) have  minimum distance $d \geq 2 \delta$ and lengths up to $q^2+q$ when $(r-1)t \leq \delta$, with $2 \leq t \leq q$. Remark \ref{cp1} and Table \ref{tabl2}, however, imply that known optimal $(r,\delta)$-LRCs with superlinear length and minimum distance $d \geq 2 \delta$ must have $r \geq (d-\delta) \geq \delta$. Hence, our optimal $(r,\delta)$-LRCs codes have new parameters. Their values for $r$ are smaller in comparison to those of prior codes. This says that erasure nodes in our coding scheme can be repaired by consulting fewer nodes.

Remark \ref{cp2} and Table \ref{tabl1} affirm that the length $n$ of previously-known optimal $(r,\delta)$-LRCs with unbounded length is \emph{divisible} by $(r+\delta-1)$. In fact, all known optimal $(r,\delta)$-LRCs in the literature have length $n$ which is divisible by $(r+\delta-1)$, except for the optimal cyclic $(r,\delta)$-LRCs over $\Ff_q$ of length $n \mid (q-1)$ or $n \mid (q+1)$ in \cite{QiuLRC}. The third family of optimal $(r,\delta)$-LRCs over $\Ff_q$ in Theorem~\ref{thm6} that we propose here is new, since the lengths are unbounded and \emph{not divisible} by $(r+\delta-1)$. \end{itemize}

In terms of organization, we proceed as follows. In Section 2, we recall some basic results which will be needed in our discussion. Section 3 is devoted to constructions of optimal $(r,\delta)$-LRCs. We make concluding remarks in Section 4 to wrap the paper up. All computations are done in {\tt MAGMA} \cite{BJP1997}.

\section{Preliminaries}\label{sec:prelims}

This section collects basic definitions and known results that will be used in our constructions. We start with the generalized Reed-Solomon codes.

\subsection{Generalized Reed-Solomon codes}\label{subsec:GRS}

Let $\Ff_q^*$ denote the multiplicative group $\Ff_q \setminus \{0\}$ of $\Ff_q$. Let $\alpha_1,\ldots,\alpha_n$ be $n$ distinct elements of $\Ff_q$. We can then define the vector $\mathbf{a}:=(\alpha_1,\ldots,\alpha_n)$. For $0 \leq k \leq n$, the generalized Reed-Solomon (GRS) code $GRS_k(\mathbf{a},\mathbf{v})$ is defined as
\begin{equation}\label{eq:GRS}
GRS_k(\mathbf{a},\mathbf{v}) := \left\{(v_1f(\alpha_1),\ldots,v_nf(\alpha_n)) : f(x) \in\Ff_q[x], \ \deg(f(x)) < k \right\},
\end{equation}
where $\mathbf{v}=(v_1,\ldots,v_n) \in (\Ff_q^*)^n$. It is well known, {\it e.g.}, from \cite[Section 9]{Coding}, that $GRS_k(\mathbf{a},\mathbf{v})$ is an $[n,k,n-k+1]_q$-MDS code. Its dual code is
\[
GRS_k(\mathbf{a},\mathbf{v})^{\perp} := GRS_{n-k}(\mathbf{a},\mathbf{v}^{\prime}),
\]
for some $\mathbf{v}^{\prime} = (v_1^{\prime},v_2^{\prime}, \ldots, v_n^{\prime})$, where $v_i^{\prime} \neq 0$ for all $1\leq i\leq n$. A parity-check matrix of $GRS_k(\mathbf{a},\mathbf{v})$ is given by
\[
H=\left(\begin{matrix}
v_1^{\prime} & v_2^{\prime} &\cdots & v_n^{\prime}&\\
v_1^{\prime}\alpha_1 & v_2^{\prime}\alpha_2 &\cdots & v_n^{\prime}\alpha_n&\\
\vdots& \vdots &  \ddots & \vdots&\\
v_1^{\prime}\alpha_1^{n-k-1} & v_2^{\prime}\alpha_2^{n-k-1} &\cdots & v_n^{\prime}\alpha_n^{n-k-1}&\\
\end{matrix}\right).
\]

One can define the \emph{extended GRS code} $GRS_k(\mathbf{a},\mathbf{v},\infty)$ of length $n+1$ to be
\[
GRS_k(\mathbf{a},\mathbf{v},\infty) := \left\{(v_1f(\alpha_1),\,\ldots, \, v_n f(\alpha_n),f_{k-1}) \, : \, f(x) \in \Ff_q[x], \ \deg(f(x)) < k \right\},
\]
where $f_{k-1}$ is the coefficient of $x^{k-1}$ in $f(x)$ and $\mathbf{v}=(v_1,\ldots,v_n)\in(\Ff_q^*)^n$. The extended GRS code preserves the MDS property and $GRS_k(\mathbf{a},\mathbf{v},\infty)$ is an $[n+1,k,n-k+2]_q$-MDS code.

\subsection{Matrix-product codes}\label{subsec:mpcodes}

 Let $A=\left(a_{i,j}\right)_{i\in[M],\,j\in[N]}$ be an $M\times N$ matrix over $\Ff_q$, with $M \leq N$. For each $i\in[M]$, let $\cc_i$ be an $[n,k_i,d_i]_q$-code. The matrix-product code $(\cc_1,\ldots,\cc_M)\cdot A$ is defined by
\begin{align}\label{MPC}
(\cc_1,\ldots,\cc_M) \cdot A
:&= \left\{(\cF_1,\ldots,\cF_M) \cdot A \, :\, \cF_1\in\cc_1, \, \ldots, \, \cF_M \in \cc_M \right\} \notag \\
&= \left\{\left(\sum_{\ell=1}^M \cF_\ell \,
a_{\ell,1},\, \ldots, \, \sum_{\ell=1}^M \cF_\ell \, a_{\ell,N}\right) \, : \, \cF_1\in\cc_1, \, \ldots, \, \cF_M \in \cc_M\right\}.
\end{align}
It is easy to check that the matrix-product code $(\cc_1,\ldots,\cc_M)\cdot A$ is a linear code of length $Nn$ over $\Ff_q$ with a generator matrix
\begin{equation}\label{eq:GenMat}
G=\left(\begin{matrix}
a_{1,1} \ G_1 & a_{1,2} \ G_1 & \cdots & a_{1,N} \ G_1\\
a_{2,1} \ G_2 & a_{2,2} \ G_2 & \cdots & a_{2,N} \ G_2\\
\vdots&\vdots &\ddots &\vdots\\
a_{M,1} \ G_M  & a_{M,2} \ G_M  & \cdots & a_{M,N} \ G_M \\
\end{matrix}
\right),
\end{equation}
where $G_i$ is a generator matrix of $\cc_i$ for $i \in[M]$.

Given an $M\times N$ matrix $A$ over $\Ff_q$, let $A_i$ stand for the $i\times N$ matrix formed by the first $i$ rows of $A$. For $1\leq j_1<j_2<\ldots<j_i\leq N$, we denote by $A(j_1,\ldots,j_i)$ the $i\times i$ matrix consisting of columns $j_1,\ldots,j_i$ of $A_i$. We call the matrix $A$ \emph{nonsingular by columns} (NSC) if $A(j_1,\ldots,j_i)$ is nonsingular for any $i\in[M]$ and $1\leq j_1<j_2<\ldots<j_i\leq N$.

\begin{example}
Let $\{\alpha_1,\ldots,\alpha_N\}$ be an arbitrary subset of $\Ff_q$. For $1\leq M\leq N\leq q$, the Vandermonde matrix
\[
\left(\begin{matrix}
1 & 1 & \cdots & 1\\
\alpha_1 & \alpha_2 & \cdots & \alpha_N\\
\vdots&\vdots &\ddots &\vdots\\
\alpha_1^{M-1}  & \alpha_2^{M-1} & \cdots & \alpha_N^{M-1} \\
\end{matrix}
\right)
\]
is nonsingular by column (NSC).
\end{example}

For an $M\times N$ NSC matrix $A$, there is no limitation on the numbers of columns if $M=1$. For $M>1$, however, it was shown in \cite{MPC1} that there exists an $M\times N$ NSC matrix over $\Ff_q$ if and only if $M\leq N\leq q$.

The next lemma gives the dimension and minimum distance of $(\cc_1,\ldots,\cc_M)\cdot A$.

\begin{lem}{\rm \cite{MPC1,MPC2}}\label{lemma21}
Let $(\cc_1,\ldots,\cc_M)\cdot A$ be a matrix-product code defined in (\ref{MPC}) with parameters $[Nn,k,d]_q$. If $A$ is full rank, then
\[
k=\sum_{i=1}^M k_i \mbox{ and }
d \geq \min_{i \in [M]} \{d_i\rho_i\},
\]
where $\rho_i$ is the minimum distance of the linear code over $\Ff_q$ generated by the first $i$ rows of $A$. Equality holds if $\cc_1,\ldots,\cc_M$ are nested codes, that is, $\cc_M\subseteq \cdots \subseteq\cc_1$. Furthermore, $\rho_i=N-i+1$, for each $i\in[M]$, if the matrix $A$ is NSC.
\end{lem}

\section{Main results}

In this section we investigate the locality of the matrix-product code $(\cc_1,\cdots,\cc_M)\cdot A$ when $\cc_1,\cdots,\cc_M$ are nested and $A$ is full rank. We establish a sufficient condition for the matrix-product code to be an optimal LRC. We then propose some new infinite families of optimal $q$-ary $(r,\delta)$-LRCs of length $n>q$. We begin with the locality of the matrix-product code $(\cc_1,\cdots,\cc_M) \cdot A$.

Suppose that $H=\left(\mathbf{h}_1,\mathbf{h}_2,\cdots,\mathbf{h}_n\right)$ is an $m\times n$ matrix. The \emph{support} of $H$ is
\[
{\rm Supp}(H) := \{i\in[n] \ : \ \mathbf{h}_i \neq \mathbf{0}\},
\]
where $\mathbf{0}$ is the zero column vector of length $m$. We will use the next lemma to determine the locality of the matrix-product code.

\begin{lem}\label{lemma1}
Let $\cc$ be an $[n,k,d]_q$-code. Let $r$ and $\delta$ be positive integers, with $\delta > 1$. The $\alpha^{\rm th}$ code symbol of $\cc$ has $(r,\delta)$-locality if and only if there exists an $m\times n$ matrix $H$ over $\Ff_{q}$ with the following properties.
\begin{enumerate}
    \item The index $\alpha$ is in ${\rm Supp}(H)$.
    \item $|{\rm Supp}(H)|\leq r+\delta-1$ such that any $\delta-1$ nonzero columns of $H$ are linearly independent over $\Ff_{q}$.
    \item $H\mathbf{c}^{\top} = \mathbf{0}$ for every $\mathbf{c}\in\cc$.
\end{enumerate}
\end{lem}

\begin{IEEEproof}
Suppose that the $\alpha^{\rm th}$ code symbol of $\cc$ has $(r,\delta)$-locality and $S_\alpha$ is a subset of $[n]$ such that $\alpha \in S_\alpha$. We let $\ell:= |S_\alpha|$ and write $S_\alpha = \{j_1,j_2,\ldots,j_\ell\}$, with $1 \leq j_1 <j_2 < \ldots <j_\ell \leq n$. Using $S_\alpha$ we construct the punctured code $\cc|_{S_\alpha}$ of length
$\ell \leq r+ \delta-1$ and distance $d\left(\cc|_{S_\alpha}\right) \geq \delta$. In deriving $\cc|_{S_\alpha}$ from $\cc$ we keep only the entries at the coordinates in $S_\alpha$, that is, we remove entries at the coordinates in $[n] \setminus S_\alpha$. Let $\overline{H} :=(\mathbf{h}_{j_1},\mathbf{h}_{j_2},\ldots,\mathbf{h}_{j_\ell})$ be a parity check matrix for $\cc|_{S_\alpha}$. By inserting $n-\ell$ zero columns into the matrix $\overline{H}$ at the coordinates in $[n] \setminus S_{\alpha}$, we obtain the matrix $H=\left(\mathbf{h}_1,\mathbf{h}_2,\cdots,\mathbf{h}_n\right)$ over $\Ff_q$ such that $\alpha\in{\rm Supp}(H)$ and $|{\rm Supp}(H)|\leq r+\delta-1$. It is immediate that $H \mathbf{c}^{\top} = \mathbf{0}$ for every $\mathbf{c} \in \cc$. Since $d\left(\cc|_{S_\alpha}\right) \geq \delta$, we can establish the $\Ff_q$-linear independence of any $\delta-1$ nonzero columns of $H$. The desired result follows.

Conversely, let $H=(h_{i,j})_{m\times n}$ be a matrix over $\Ff_{q}$ that has the three listed properties. Let $\cc|_{{\rm Supp}(H)}$ be the code derived from $\cc$ by puncturing at the coordinates in $[n] \setminus {\rm Supp}(H)$. We write $H$ in terms of its $n$ column vectors as $H=(\mathbf{h}_1,\mathbf{h}_2, \ldots, \mathbf{h}_n)$. Then $K:=(\mathbf{h}_j)_{j\in{\rm Supp}(H)}$ is an $m\times |{\rm Supp}(H)|$ matrix such that $K \mathbf{a}^{\top} = \mathbf{0}$ for each $\mathbf{a}\in\cc|_{{\rm Supp}(H)}$. Since any $\delta-1$ columns of $K$ are $\Ff_{q}$-linearly independent, the minimum distance of $\cc|_{{\rm Supp}(H)}$ is at least $\delta$. Thus, for any $\alpha\in{\rm Supp}(H)$, the $\alpha^{\rm th}$ code symbol of $\cc$ has $(r,\delta)$-locality.
\end{IEEEproof}

We now characterize the locality of the matrix-product code $(\cc_1,\ldots,\cc_M)\cdot A$ in terms of the locality of $\cc_1$, given that $\cc_1,\ldots,\cc_M$ are nested codes.

\begin{lem}\label{lemma2}
Let $\cc_1,\ldots,\cc_M$ be $q$-ary linear codes of length $n$ with $\cc_M \subseteq \cdots \subseteq \cc_1$. Let $\cc:=(\cc_1,\ldots,\cc_M) \cdot A$ be the matrix-product code over $\Ff_q$ defined in (\ref{MPC}), where $A$ has full rank. If $\cc_1$ is an $(r,\delta)$-LRC, then $\cc$ has $(r,\delta)$-locality.
\end{lem}

\begin{IEEEproof}
By Lemma \ref{lemma1} and the fact that $\cc_1$ is an $(r,\delta)$-LRC, for any $\alpha\in[n]$, there exists an $m \times n$ matrix $H$ over $\Ff_{q}$ with $\alpha\in{\rm Supp}(H)$, the support size $|{\rm Supp}(H)| \leq r+ \delta-1$ such that any $\delta-1$ nonzero columns of $H$ are linearly independent over $\Ff_{q}$, and $H\mathbf{c}^{\top}=\mathbf{0}$ for every $\mathbf{c}\in\cc_1$.

Let $O$ denote the $m\times n$ zero matrix and let $\mathcal{H}_1 := \left(H,O,\cdots,O\right)$. The latter is an $m\times nN$ matrix over $\Ff_{q}$ that inherits the three relevant properties of $H$, namely, $\alpha\in{\rm Supp}(\mathcal{H}_1)$, the support size $|{\rm Supp}(\mathcal{H}_1)| \leq r+\delta-1$, and any $\delta-1$ nonzero columns of $\mathcal{H}_1$ are linearly independent over $\Ff_{q}$.
Note that
\[
\cc=(\cc_1,\ldots,\cc_M)\cdot A=\left\{\left(\sum_{\ell=1}^M\cF_\ell a_{\ell1},\cdots,\sum_{\ell=1}^M\cF_\ell a_{\ell N}\right) \, : \, \cF_1 \in \cc_1, \ldots, \cF_M \in \cc_M \right\}.
\]
It follows from $\cc_M\subseteq \cdots \subseteq \cc_1$ that $\mathcal{H}_1\mathbf{c}^{\top}=\mathbf{0}$ for any $\mathbf{c}\in\cc$. By Lemma \ref{lemma1}, the $\alpha^{\rm th}$ symbol of $\cc$ has $(r,\delta)$-locality for any $\alpha\in[n]$. Continuing inductively, by defining
\begin{eqnarray*}
\mathcal{H}_2:=\left(O,H,\cdots,O\right),\, \ldots, \, \mathcal{H}_N:=\left(O,O,\cdots,H\right),
\end{eqnarray*}
we can similarly infer that the $(jn+\alpha)^{\rm th}$ symbol of $\cc$ has $(r,\delta)$-locality for each $j \in[N-1]$. Thus, $\cc$ is an $(r,\delta)$-LRC.
\end{IEEEproof}

\subsection{Optimal $(r,\delta)$-LRCs with $d\geq 2\delta$}

We now look into matrix-product codes from linear MDS codes. We present an easily computable criterion for a matrix-product code to be optimal. %with respect to the bound in (\ref{LRCbound}).
We then construct optimal $(r,\delta)$-LRCs over $\Ff_q$ of lengths up to $q^2+q$. To establish the criterion we use the next lemma.

\begin{lem}\label{lemd}
Let $r$, $\delta$, $M$ and $N$ be positive integers, with $\delta>1$ and $q\geq N>M>1$. Let $\cc_1 = \cdots = \cc_{M-1}$ be $[r+\delta-1,r,\delta]_q$-MDS codes. Let $\cc_M$ be an $[r+\delta-1,r-g,\delta+g]_q$- MDS code, where $g<r$ is a nonnegative integer. Let $\cc=(\cc_1,\ldots,\cc_M) \cdot A$, as defined in (\ref{MPC}), be a matrix-product code such that $A$ is an $M\times N$ NSC matrix. If
\[
\cc_M \subseteq \cc_1, \quad 0 \leq g \leq \frac{\delta}{N-M+1} \mbox{, and } (N-M)(g-r+1) =\left\lfloor \frac{g}{r} \right\rfloor(\delta-1),
\]
then $\cc$ is an optimal $(r,\delta)$-LRC with parameters
\[
\left[N(r+\delta-1),Mr-g,(N-M+1)(\delta+g)\right]_q.
\]
\end{lem}

\begin{IEEEproof}
It is straightforward to confirm that $\cc_1$ has $(r,\delta)$-locality. Since  $\cc_M \subseteq \cdots \subseteq \cc_1$ and $A$ is NSC, it follows from Lemma \ref{lemma2} that $\cc=(\cc_1,\ldots,\cc_M)\cdot A$ is an $(r,\delta)$-LRC. Therefore, by Lemma \ref{lemma21}, $\cc$ has parameters
\[
[Nn,Mr-g,d]_q \mbox{, with }
d={\rm min} \left\{N \delta,\, \ldots,\, (N-M+2) \delta, \, (N-M+1)(\delta+g)\right\}.
\]
Since $0\leq g\leq \frac{\delta}{N-M+1}$, the minimum distance $d$ is $(N-M+1) (\delta+g)$. By the bound in (\ref{LRCbound}), we get the upper bound
\begin{align}
d &\leq N(r+\delta-1)-(Mr-g)- \left(\left\lceil \frac{Mr-g}{r} \right\rceil-1\right)(\delta-1)+1 \notag \\
&= (N-M)(r+\delta-1) +g + \delta+ \left\lfloor \frac{g}{r} \right\rfloor(\delta-1) \notag\\
&= (N-M+1)(\delta+g).
\end{align}
The last equality comes from $(N-M)(g-r+1) = \left\lfloor \frac{g}{r} \right\rfloor(\delta-1)$. Thus, $\cc$ is an optimal $(r,\delta)$-LRC.
\end{IEEEproof}

Lemma \ref{lemd} leads to a nice characterization.

\begin{thm}\label{thm1}
Let $r$, $\delta$, $M$, and $N$ be positive integers, with $\delta>1$ and $q \geq N > M >1$. Let $\cc_1=\cdots=\cc_{M-1}$ be $[r+\delta-1,r,\delta]_q$-MDS codes. Let $g$ be an integer with
\[
0\leq g\leq \frac{\delta}{N-M+1} \mbox{ and } g<r.
\]
Let $\cc_M$ be an $[r+\delta-1,r-g,\delta+g]_q$-MDS code such that $\cc_M \subseteq \cc_1$. Suppose that $\cc=(\cc_1,\cdots,\cc_M)\cdot A$, as defined in (\ref{MPC}), is the matrix-product code such that $A$ is an $M\times N$ NSC matrix. Then $\cc$ is an optimal $(r,\delta)$-LRC if and only if $r=g+1$.
\end{thm}

\begin{IEEEproof}
If $r=g+1$, then the assertion follows immediately from Lemma \ref{lemd}.

Conversely, by Lemma \ref{lemd}, the matrix-product code $\cc$ has parameters
\[
\left[N(r+\delta-1),Mr-g,(N-M+1)(\delta+g)\right]_q.
\]
Since $\cc$ is optimal, we obtain
\[
(N-M)(g-r+1)=\left\lfloor\frac{g}{r}\right\rfloor(\delta-1).
\]
Since $r-g>0$ and $N>M$, we have $r=g+1$.
\end{IEEEproof}

Theorem \ref{thm1} describes a method to use nested MDS codes in generating optimal LRCs. It is well known, {\it e.g.}, from \cite[Chapter 7, Section 4]{Huff}, that trivial $q$-ary linear MDS codes are the repetition codes $[\ell,1,\ell]_q$, their duals $[\ell,\ell-1,2]_q$, and the full vector space $[\ell,\ell,1]_q$ of any length $\ell$. The main conjecture for MDS codes says that a nontrivial $q$-ary MDS code of length $\ell$ and dimension $k$ exists if $\ell\leq q+1$, except when $q$ is even and $k=3$ or $k=q-1$, in which case $\ell\leq q+2$.

Numerous explicit constructions of $[\ell,k,\ell-k+1]_q$-MDS codes for any $\ell \in [q+1]$ and $k \in [q]$ are known. For our current purpose, it suffices to use cyclic, constacyclic, GRS, and extended GRS constructions of MDS codes (see, {\it e.g.}, \cite[Chapters 4 and 5]{Huff}). The GRS codes associated with the same vectors $\mathbf{a}$ and $\mathbf{v}$ are nested, by definition. Using $q$-ary GRS codes in Theorem \ref{thm1} yields the following corollary.

\begin{cor}\label{cor1}
Let the notation be as in Theorem \ref{thm1}. Let $\cc_1=\cdots=\cc_{M-1}=GRS_r(\mathbf{a},\mathbf{v})$ with parameters $[r+\delta-1,r,\delta]_q$ and $\cc_M=GRS_{1}(\mathbf{a},\mathbf{v})$. If $ (r-1)(N-M+1)\leq \delta$, then the matrix-product code $\cc$ is an optimal $(r,\delta)$-LRC with parameters
\begin{equation}\label{eq:cor1}
\left[N(r+\delta-1),(M-1)r+1,(N-M+1)(r+\delta-1)\right]_q.
\end{equation}
\end{cor}

%We give an example to illustrate Corollary \ref{cor1}.

\begin{example}
Let $\cc_1 :=GRS_2((1,3,2,6,4),(1,1,1,1,1))$ be defined over $\Ff_7$. This code has parameters $[5,2,4]_7$. Over the same field, let $\cc_2:=GRS_{1}((1,3,2,6,4),(1,1,1,1,1))$. Notice that $\cc_2$ is a subcode of $\cc_1$. Let
\[
A=\begin{pmatrix}
1 & 1 & 1\\
1 & 2 & 3
\end{pmatrix}
\]
be a matrix over $\Ff_7$. The matrix-product code $\cc :=(\cc_1,\cc_2) \cdot A$ is a $[15,3,10]_7$-code with a generator matrix
\[
A=\begin{pmat}({....|....|.....})
1 & 1 & 1 & 1 &1 & 1 & 1 & 1 & 1 &1 & 1 & 1 & 1 & 1 &1\cr
1 & 3 & 2 & 6 & 4 & 1 & 3 & 2 & 6 & 4 & 1 & 3 & 2 & 6 & 4\cr\-
1 & 1 & 1 & 1 &1 & 2 & 2 & 2 & 2 &2 & 3 & 3 & 3 & 3 &3\cr
\end{pmat}.
\]
Each of its punctured codes $\cc|_{\{1,2,3,4,5\}}$, $\cc|_{\{6,7,8,9,10\}}$, and $\cc|_{\{11,12,13,14,15\}}$ has parameters $[5,2,4]_7$. By Corollary \ref{cor1}, $\cc$ is an optimal $7$-ary $(2,4)$-LRC.
\end{example}

The optimal $(r,\delta)$-LRCs constructed by Corollary \ref{cor1} have flexible parameters. By choosing $q=5$, for instance, one can determine that the total number of optimal $(r,\delta)$-LRCs generated by Corollary \ref{cor1} is $38$. Table \ref{table3} lists all of these optimal $(r,\delta)$-LRCs. Asymptotically, the total number of optimal $(r,\delta)$-LRCs produced by Corollary \ref{cor1} increases in a superlinear manner with respect to $q$.

\begin{table}[!htbp]
\caption{The parameters $[n,k,d]$ of optimal $(r,\delta)$-LRCs generated by Corollary \ref{cor1} for $q=5$}
\label{table3}
\renewcommand{\arraystretch}{1.2}
\centering
 \begin{tabular}{c ccccl | c ccccl}
\toprule
No. & $N$ & $M$ & $r$ & $\delta$ & $[n,k,d]$ & No. & $N$ & $M$ & $r$ & $\delta$ & $[n,k,d]$ \\
\midrule
$1$ & $3$&	$2$&$1$&	$2$&	$[6,	2,	4]$&
$20$ & $4$&	$3$&	$2$&	$2$&	$[12,	5,	6]$\\

$2$ & &	&	$2$&	$2$&	$[9,	3,	6]$&
$21$ &&&	$1$&	$3$&	$[12,	3,	6]$\\

$3$ & &	&	$1$&	$3$&	$[9,	2,	6]$&
$22$ & &&	$2$&	$3$&	$[16,	5,	8]$\\

$4$ & &&	$2$&	$3$&	$[12,	3,	8]$&
$23$ & &&	$1$&	$4$&	$[16,	3,	8]$\\

$5$ & &&	$1$&	$4$&	$[12,	2,	8]$&
$24$ & &&	$2$&	$4$&	$[20,	5,	10]$\\

$6$ & &&	$2$&	$4$&	$[15,	3,	10]$&
$25$ &&&	$1$&	$5$&	$[20,	3,	10]$\\

$7$ & &&	$1$&	$5$&	$[15,	2,	10]$&
$26$ & $5$&	$3$&	$1$&	$2$&	$[10,3,6]$\\

$8$ & $4$&	$2$&	$1$&	$2$&	$[8,2,6]$&
$27$ & &&	$1$&	$3$&	$[15,	3,	9]$\\

$9$ &&&	$1$&	$3$&	$[12,	2,	9]$&
$28$ &&&	$2$&	$3$&	$[20,	5,	12]$\\

$10$ &&&	$2$&	$3$&	$[16,	3,	12]$&
$29$ &&&	$1$&	$4$&	$[20,	3,	12]$\\

$11$ &&&	$1$&	$4$&	$[16,	2,	12]$&
$30$ &&&	$2$&	$4$&	$[25,	5,	15]$\\

$12$ &&&	$2$&	$4$&	$[20,	3,	15]$&
$31$ &&&	$1$&	$5$&	$[25,	3,	15]$\\

$13$ &&&	$1$&	$5$&	$[20,	2,	15]$&
$32$ & $5$&	$4$&	$1$&	$2$&	$[10,4,	4]$\\
$14$ &$5$&	$2$&	$1$&$2$&	$[10,	2,	8]$& $33$ &&&	$2$&	$2$&	$[15,	7,	6]$\\

$15$ & &&	$1$&	$3$&	$[15,	2,	12]$&
$34$ &&&	$1$&	$3$&	$[15,	4,	6]$\\

$16$ & &&	$1$&	$4$&	$[20,	2,	16]$&
$35$ &&&	$2$&	$3$&	$[20,	7,	8]$\\

$17$ &&&	$2$&	$4$&	$[25,	3,	20]$&
$36$ &&&	$1$&	$4$&	$[20,	4,	8]$\\

$18$ &&&	$1$&	$5$&	$[25,	2,	20]$&
$37$ &&&	$2$&	$4$&	$[25,	7,	10]$\\

$19$ &$4$&	$3$&	$1$&	$2$&	$[8,3,	4]$&
$38$ &&&	$1$&	$5$&	$[25,	4,	10]$\\
\bottomrule
\end{tabular}
\end{table}

In Corollary \ref{cor1}, we deploy $q$-ary MDS codes of length $\ell \leq q$ to design optimal $(r,\delta)$-LRCs. Now we utilize $q$-ary MDS codes of length $q+1$. Let us denote by $\cc_{k}^{q+1}$ an MDS code of length $q+1$ and dimension $k$. It was shown in \cite{MDS1} that nested MDS codes $\cc_{k}^{q+1}\subseteq\cc_{k+1}^{q+1}$ over $\Ff_q$ exist only when $q$ is even and $k=2$ or $k=q-2$. Such nested codes are not suitable as construction ingredients in Theorem \ref{thm1}. Ezerman, Grassl, and Sol\'{e} in \cite{MDS2} presented nested MDS codes of length $q+1$ and codimension $2$ as follows.

\begin{lem}\label{lemmds}
For $k \in [q+1]$, let $\cc_{k}^{q+1}$ be the cyclic or constacyclic $[q+1,k,q-k+2]_q$-MDS code constructed in the proof of {\rm \cite[Theorem~8]{MDS2}. By how the construction works,} the codes of odd dimension and the codes of even dimension form two sequences of nested MDS codes, that is,
\begin{align}
&\cc_{1}^{q+1} \subset \cc_{3}^{q+1} \subset \cdots
\subset \cc_{j}^{q+1} \mbox{, where } j \mbox{ is odd and } (q+1-j) \leq 1, \label{eq:oddnest}\\
&\cc_{2}^{q+1} \subset \cc_{4}^{q+1} \subset \cdots
\subset \cc_{j}^{q+1} \mbox{, where } j \mbox{ is even and } (q+1-j) \leq 1. \label{evennest}
\end{align}
\end{lem}

From \cite[Section V]{MDS2} we know that any $q$-ary MDS code of length $q+1$ has a codeword of weight $q+1$, except for the $q$-ary simplex code with parameters $[q+1,2,q]_q$ that has only codewords of weights $0$ and $q$. Hence, by Lemma \ref{lemmds} and Theorem \ref{thm1}, we have the following construction of optimal $(r,\delta)$-LRCs.

\begin{cor}\label{cor2}
Let the notation be as in Theorem \ref{thm1}. Let $\cc_{1}^{q+1}\subseteq\cc_{r}^{q+1}$ be nested MDS codes given in Lemma \ref{lemmds} as (\ref{eq:oddnest}), where $r\geq 1$ is odd. Assume that $\cc_1=\cdots=\cc_{M-1}=\cc_{r}^{q+1}$ have parameters $[q+1,r,q+2-r]_q$ and $\cc_M=\cc_{1}^{q+1}$ has parameters $[q+1,1,q+1]_q$. If $(r-1)(N-M+1)+r\leq q+2$, then the matrix-product code $\cc$ is an optimal $(r,q+2-r)$-LRC with parameters
$\left[N(q+1),(M-1)r+1,(N-M+1)(q+1)\right]_q$.
\end{cor}

%Corollary \ref{cor2} can be illustrated briefly by the following instance.

\begin{example}
Let $\alpha \in \Ff_{25}$ be a primitive
$6^{\rm th}$ root of unity. Let $\cc_1=\cc_2=\cc_3$ be the $[6,3,4]_5$-cyclic code with generator polynomial
\[
g(x)= \left(x-\alpha^2\right) \left(x-\alpha^3\right) \left(x-\alpha^4\right) = x^3+2x^2+2x+1
\]
and let $\cc_4$ be the $[6,1,6]_5$-cyclic code with generator polynomial
\[
f(x)=(x-\alpha)(x-\alpha^2)(x-\alpha^3)(x-\alpha^4)(x-\alpha^5)=x^5+x^4+x^3+x^2+x+1.
\]
The matrix
\[
A=\begin{pmatrix}
1 & 1 & 1 & 1 & 1\\
0 & 1 & 2 & 3 &4 \\
0 & 1 & 2^2 & 3^2 & 4^2\\
0& 1 & 2^3 & 3^3 & 4^3
\end{pmatrix}
\]
is NSC over $\Ff_5$. The matrix-product code $\cc=(\cc_1,\cc_2,\cc_3,\cc_4)\cdot A$ has parameters $[30,10,12]_5$ and a generator matrix
\[
G=\begin{pmat}({.....|.....|.....|.....|......})
1 & 2 & 2 & 1 & 0 & 0 & 1 & 2 & 2 & 1 & 0 & 0 & 1 & 2 & 2 & 1 & 0 & 0 & 1 & 2 & 2 & 1 & 0 & 0 & 1 & 2 & 2 & 1 & 0 & 0 \cr
0 & 1 & 2 & 2 & 1 & 0 & 0 & 1 & 2 & 2 & 1 & 0 & 0 & 1 & 2 & 2 & 1 & 0 & 0 & 1 & 2 & 2 & 1 & 0 & 0 & 1 & 2 & 2 & 1 & 0 \cr
0 & 0 & 1 & 2 & 2 & 1 & 0 & 0 & 1 & 2 & 2 & 1 & 0 & 0 & 1 & 2 & 2 & 1 & 0 & 0 & 1 & 2 & 2 & 1 & 0 & 0 & 1 & 2 & 2 & 1 \cr\-
0 & 0 & 0 & 0 & 0 & 0 & 1 & 2 & 2 & 1 & 0 & 0 & 2 & 4 & 4 & 2 & 0 & 0 & 3 & 1 & 1 & 3 & 0 & 0 & 4 & 3 & 3 & 4 & 0 & 0 \cr
0 & 0 & 0 & 0 & 0 & 0 & 0 & 1 & 2 & 2 & 1 & 0 & 0 & 2 & 4 & 4 & 2 & 0 & 0 & 3 & 1 & 1 & 3 & 0 & 0 & 4 & 3 & 3 & 4 & 0 \cr
0 & 0 & 0 & 0 & 0 & 0 & 0 & 0 & 1 & 2 & 2 & 1 & 0 & 0 & 2 & 4 & 4 & 2 & 0 & 0 & 3 & 1 & 1 & 3 & 0 & 0 & 4 & 3 & 3 & 4 \cr\-
0 & 0 & 0 & 0 & 0 & 0 & 1 & 2 & 2 & 1 & 0 & 0 & 4 & 3 & 3 & 4 & 0 & 0 & 4 & 3 & 3 & 4 & 0 & 0 & 1 & 2 & 2 & 1 & 0 & 0 \cr
0 & 0 & 0 & 0 & 0 & 0 & 0 & 1 & 2 & 2 & 1 & 0 & 0 & 4 & 3 & 3 & 4 & 0 & 0 & 4 & 3 & 3 & 4 & 0 & 0 & 1 & 2 & 2 & 1 & 0 \cr
0 & 0 & 0 & 0 & 0 & 0 & 0 & 0 & 1 & 2 & 2 & 1 & 0 & 0 & 4 & 3 & 3 & 4 & 0 & 0 & 4 & 3 & 3 & 4 & 0 & 0 & 1 & 2 & 2 & 1\cr\-
0 & 0 & 0 & 0 & 0 & 0 & 1 & 1 & 1 & 1 & 1 & 1 & 3 & 3 & 3 & 3 & 3 & 3 & 2 & 2 & 2 & 2 & 2 & 2 & 4 & 4 & 4 & 4 & 4 & 4 \cr
\end{pmat}.
\]
Its punctured codes
\[
\cc|_{\{1,2,3,4,5,6\}}, \quad \cc|_{\{7,8,9,10,11,12\}}, \quad \cc|_{\{13,14,15,16,17,18\}}, \quad \cc|_{\{19,20,21,22,23,24\}}, \mbox{ and }  \cc|_{\{25,26,27,28,29,30\}}
\]
have parameters $[6,3,4]_5$. The code $\cc$ is an optimal $5$-ary $(3,4)$-LRC, by Corollary \ref{cor2}.
\end{example}

\begin{remark}
An optimal $(r,\delta)$-LRC produced by Corollary \ref{cor1} has length up to $q^2$. The maximum length of an optimal $(r,\delta)$-LRC generated by Corollary \ref{cor2} is $q^2+q$. Thus, Corollaries \ref{cor1} and \ref{cor2} allow us to construct optimal LRCs of superlinear lengths.
\end{remark}

\begin{remark}\label{cp1}
To highlight that the parameters of optimal $(r,\delta)$-LRCs constructed by Corollaries \ref{cor1} and \ref{cor2} are new, we list the parameters of known optimal $(r,\delta)$-LRCs with superlinear length and minimum distance $d \geq 2 \delta$ alongside our results in Table \ref{tabl2}. The table shows that all prior optimal $(r,\delta)$-LRCs in the literature satisfy the condition $r \geq (d-\delta) > \delta$. The optimal $(r,\delta)$-LRCs produced by Corollaries \ref{cor1} and \ref{cor2}, on the other hand, meet the condition $(r-1) \ t \leq \delta$, for $2 \leq t \leq q$. This means that our optimal $(r,\delta)$-LRCs have new parameters. Most notably, the parameter $r$ in our constructions is smaller than that in the literature. Thus, the erasure symbols can be repaired by less amount of data compared with the amount required in prior literature.
\end{remark}

\begin{table*}[!htbp]
\caption{The parameters of $q$-ary optimal $(r,\delta)$-LRCs with superlinear length $n$ and minimum distance $d\geq 2\delta$}
\label{tabl2}
\renewcommand{\arraystretch}{1.2}
\centering
\begin{tabular}{ccll}
\toprule
No.  & Distance  & Constraints & References \\ \midrule
$1$ &    $d \in \{5,6\}$ & $\delta=2, \quad r\geq d-2$, \quad $(r+1) \mbox{ divides } n$ & \cite{Cai,LRClength,LRCJin}  \\

$2$ &    $d\geq7$ & $\delta=2, \quad r\geq d-2$, \quad $(r+1) \mbox{ divides } n$ & \cite{LRCShang,LRCXing}  \\

$3$ &   $d\geq 2\delta+1$ & $ r\geq d-\delta+1$, \quad $(r+\delta-1) \mbox{ divides } n$, & \cite{Cai}  \\

$4$ &   $2\delta+1\leq d\leq 3\delta$ & $ r\geq d-\delta+1$, \quad $(r+\delta-1) \mbox{ divides } n$, & \cite{LRCZhang2}  \\

$5$ & $d\geq 3\delta+1$ & $r\geq d-\delta$, \quad $(r+\delta-1) \mbox{ divides } n$ & \cite{LRCKong,LRCZhang1} \\

$6$ & $d=t \ (r+\delta-1)$ & $2 \leq t \leq q$, \quad $(r-1) \ t \leq \delta$, \quad $(r+\delta-1) \mbox{ divides } n$ & Corollaries \ref{cor1} and \ref{cor2}  \\
\bottomrule
\end{tabular}
\end{table*}

\subsection{Optimal $(r,\delta)$-LRCs with $\delta \leq d\leq 2\delta$}

This subsection investigates the matrix-product code $(\cc_1,\cdots,\cc_M)\cdot A$, when $A$ is a square matrix. In this case, we can relax the condition that $\cc_1,\cdots,\cc_M$ must be MDS codes in Theorem \ref{thm1} to requiring that $\cc_1,\cdots,\cc_M$ be optimal $(r,\delta)$-LRCs. This allows us to propose a new construction of optimal $(r,\delta)$-LRCs of any length \emph{not divisible} by $(r+\delta-1)$.

\begin{thm}\label{thm2}
Let $r$, $\delta$, $w$, and $N$ be positive integers such that $\delta>1$ and $q\geq N>1$. Let $\cc_1=\cdots=\cc_{N-1}$ be codes with parameters $[n,k = wr-s,\delta]_q$, where $0 \leq s <r$. Let $\cc_N$ be an $[n,k-i,\delta+i]_q$-code with $0 \leq i \leq \delta$. Let $\cc=(\cc_1,\cdots,\cc_N)\cdot A$ be the matrix-product code constructed as in (\ref{MPC}) such that $A$ is an NSC matrix of order $N$. If $\cc_1$ and $\cc_N$ are optimal $(r,\delta)$-LRCs that satisfy $\cc_N \subseteq \cc_1$ and $Ns+i <r$, then $\cc$ is an optimal $(r,\delta)$-LRC with parameters $\left[Nn,Nk-i,\delta+i\right]_q$.
\end{thm}

\begin{IEEEproof}
Since $\cc_1$ has $(r,\delta)$-locality, it follows from Lemma \ref{lemma2} that $\cc$ is an $(r,\delta)$-LRC. Since $\cc_1$ is optimal, %with respect to the bound in (\ref{LRCbound}),
\[
n-k=\left\lceil\frac{k}{r}\right\rceil(\delta-1).
\]
By Lemma \ref{lemma21} and since $0 \leq i \leq \delta$, we know that $\cc$ is an $[Nn,Nk-i,\delta+i]_q$-code. Thanks to the bound in (\ref{LRCbound}) for $\cc$ and because $Ns+i<r$, we can establish that $\cc$ is optimal, since
\begin{align*}
d &\leq Nn-(Nk-i)-\left(\left\lceil\frac{Nk-i}{r}\right\rceil-1\right)(\delta-1)+1\\
&= N\left\lceil\frac{k}{r}\right\rceil(\delta-1)+i-\left(\left\lceil\frac{Nk-i}{r}\right\rceil-1\right)(\delta-1)+1\\
&= \delta+i+\left(N\left\lceil\frac{k}{r}\right\rceil-\left\lceil\frac{Nk-i}{r}\right\rceil\right)(\delta-1)\\
&=\delta+i.
\end{align*}
\end{IEEEproof}

To meet the assumptions in Theorem \ref{thm2}, one needs a sequence of nested optimal $(r,\delta)$-LRCs that satisfies certain conditions. Such nested optimal $(r,\delta)$-LRCs are in fact already discussed in \cite{Cai,ChenBC,LRC2,LRCC2,QiuLRC}. Most known optimal $(r,\delta)$-LRCs from these references have length $n$ which is a multiple of $(r+\delta-1)$. Consequently, in order to design optimal $(r,\delta)$-LRCs with new parameters, we propose the following construction of $(r,\delta)$-LRCs of lengths not divisible by $(r+\delta-1)$.

We begin with a construction of optimal $(r,\delta)$-LRCs with minimum distance $\delta$. Let $x\in\Ff_q$ and let $f_{[i,j]}(x)$ be a column vector of length $j-i+1$ defined by
\[
f_{[i,j]}(x)=(x^i,x^{i+1},\ldots,x^j)^{\top}.
\]
We build an $[m=r+\delta-1,r,\delta]_q$-GRS code $\mathcal{E}:=GRS_r(\mathbf{a},\mathbf{v})$ with a parity-check matrix
\begin{equation}\label{m1}
H=\left(\lambda_1f_{[0,\delta-2]}(\alpha_1),\cdots,\lambda_mf_{[0,\delta-2]}(\alpha_m)\right).
\end{equation}
Let $\mathbf{0}_{\delta-1}$ be the zero column vector of length $\delta-1$. We use
\begin{align*}
H_{1} &=\begin{pmatrix}
\lambda_1f_{[0,\delta-2]}(\alpha_1) & \cdots &  \lambda_{m-1}f_{[0,\delta-2]}(\alpha_{m-1})\\
\mathbf{0}_{\delta-1} & \cdots & \mathbf{0}_{\delta-1}
\end{pmatrix}
\mbox{ and}\\
H_{2} &=\begin{pmatrix}
\lambda_{m}f_{[0,\delta-2]}(\alpha_{m}) & \mathbf{0}_{\delta-1} & \cdots & \mathbf{0}_{\delta-1}\\
\lambda_{m}f_{[0,\delta-2]}(\alpha_{m}) & \lambda_{m-1}f_{[0,\delta-2]}(\alpha_{m-1}) & \cdots &  \lambda_{1}f_{[0,\delta-2]}(\alpha_{1})
\end{pmatrix}
\end{align*}
to construct a $2(\delta-1)\times (2m-1)$ matrix
\begin{equation}\label{m2}
\widehat{H}:=
\begin{pmatrix}H_{1} & H_{2}\end{pmatrix}.
\end{equation}
It is clear that $\rank(\widehat{H})=2(\delta-1)$. By setting $\widehat{H}$ to be a parity-check matrix of a $q$-ary linear code, we derive the following result.

\begin{thm}\label{thm3}
Let $r$, $\delta$, and $m:= r + \delta -1 \leq q$ be positive integers. Let $\mathcal{D}$ be the $q$-ary linear code whose parity-check matrix is the matrix $\widehat{H}$ in (\ref{m2}). Then $\mathcal{D}$ is a $[2m-1,2r-1,\delta]_q$-optimal $(r,\delta)$-LRC.
\end{thm}
\begin{IEEEproof}
By how $\mathcal{D}$ is constructed, the length and dimension of $\mathcal{D}$ are $2(r+\delta-1)-1$ and $2r-1$, respectively. Since $H$ is a parity-check matrix of an $[r+\delta-1,r,\delta]_q$-GRS code, any $\delta-1$ columns of $H$ are linearly independent. For any $i \in [r+\delta-1]$, there exists an $(\delta-1)\times (2r+2\delta-3)$ matrix $\begin{pmatrix} H & O \end{pmatrix}$
over $\Ff_{q}$ with $i\in{\rm Supp}(H)$ and $|{\rm Supp}(H)|=r+\delta-1$, where $O$ is the $(\delta-1)\times(r+\delta-2)$ zero matrix. It is evident that any $\delta-1$ nonzero columns of $\begin{pmatrix} H & O \end{pmatrix}$ are $\Ff_{q}$-linearly independent and
$\begin{pmatrix} H & O \end{pmatrix} \cdot \mathbf{c}^{\top}= \mathbf{0}$ for every $\mathbf{c}\in\mathcal{D}$. It follows from Lemma \ref{lemma1} that the $i^{\rm th}$ code symbol of $\mathcal{D}$ has $(r,\delta)$-locality, for any $i\in[r+\delta-1]$.
%Using an argument similar to that above,
We can subsequently show that $\mathcal{D}$ is an $(r,\delta)$-LRC. By the bound in (\ref{LRCbound}), the minimum distance $d$ of $\mathcal{D}$ satisfies
\begin{equation*}
d \leq  2 \ (r+\delta-1)-1-(2r-1)- \left(\left\lceil\frac{2r-1}{r}\right\rceil-1\right) (\delta-1)+1 = \delta.
\end{equation*}

It remains to show that $d\geq\delta$. We do this by demonstrating that any $\delta-1$ columns of $\widehat{H}$ are $\Ff_q$-linearly independent in each of the following two cases.

\begin{enumerate}[wide, itemsep=0pt, leftmargin =0pt,widest={{\bf Case $2$}}]
\item[{\bf Case $1$}:]
We consider any $\delta-1$ columns of $\widehat{H}$, excluding
$\begin{pmatrix} \lambda_{m}f_{[0,\delta-2]}(\alpha_{m}) \\ \lambda_{m}f_{[0,\delta-2]}(\alpha_{m})
\end{pmatrix}$.
Let
\begin{equation}
\sum_{s=1}^{\mu}a_{i_s}\begin{pmatrix} \lambda_{i_s}f_{[0,\delta-2]}(\alpha_{i_s}) \\ \mathbf{0}_{\delta-1} \end{pmatrix}+\sum_{t=1}^{\nu}b_{j_t}\begin{pmatrix} \mathbf{0}_{\delta-1} \\ \lambda_{j_t}f_{[0,\delta-2]}(\alpha_{j_t}) \end{pmatrix}=\mathbf{0},
\end{equation}
where
\[
\mu+\nu=\delta-1, \quad
1\leq i_1 < \ldots <i_\mu\leq r+\delta-2, \quad
1\leq j_1<\ldots<j_\nu\leq r+\delta-2,
\]
and $a_{i_1},\ldots,a_{i_\mu},b_{j_1},\ldots,b_{j_\nu}$ are elements of $\Ff_q$. Then we have
\begin{align*}
a_{i_1} \lambda_{i_1} f_{[0,\delta-2]} (\alpha_{i_1}) + \ldots + a_{i_\mu}\lambda_{i_\mu} f_{[0,\delta-2]}(\alpha_{i_\mu}) &= \mathbf{0}_{\delta-1} \mbox{ and}\\
b_{j_1} \lambda_{i_1}f_{[0,\delta-2]}(\alpha_{j_1}) + \ldots+b_{j_\mu} \lambda_{j_\mu} f_{[0,\delta-2]}(\alpha_{j_\mu}) &=
\mathbf{0}_{\delta-1}.
\end{align*}
Since any $\delta-1$ columns of $H$ are linearly independent, we obtain $a_{i_s}=b_{j_t}=0$, for any $s\in[\mu]$ and $t\in[\nu]$.

\item[{\bf Case $2$}:]
We can argue similarly that any $\delta-1$ columns of $\widehat{H}$, including $\begin{pmatrix} \lambda_{m}f_{[0,\delta-2]}(\alpha_{m}) \\ \lambda_{m}f_{[0,\delta-2]}(\alpha_{m})
\end{pmatrix}$, are linearly independent. Thus, $d \geq \delta$, confirming that $\mathcal{D}$ is an optimal $(r,\delta)$-LRC.
\end{enumerate}
\end{IEEEproof}

\begin{example}
Let $r=2$, $\delta=3$, $q=5$, and
\[
\widehat{H}=\begin{pmatrix}
1 & 1 & 1 & 1 & 0 & 0 & 0 \\
1 & 2 & 3 & 4 & 0 & 0 & 0 \\
0 & 0 & 0 & 1 & 1 & 1 & 1\\
0 & 0 & 0 & 4 & 3 & 2 & 1
\end{pmatrix}.
\]
The code $\mathcal{D}$ with parity-check matrix $\widehat{H}$ has parameters $[7,3,3]_5$. The punctured codes $\cc|_{\{1,2,3,4\}}$ and $\cc|_{\{4,5,6,7\}}$ are of parameters $[4,2,3]_5$, making $\mathcal{D}$ an optimal $(2,3)$-LRC over $\Ff_5$.
\end{example}

Given codes $\cc_1$ and $\cc_2$ with parameters $[n_1,k_1,d_1]_q$ and $[n_2,k_2,d_2]_q$, respectively, their direct sum is
\[
\cc_1 \oplus \cc_2 := \{(\mathbf{c}_1,\mathbf{c}_2)\,:\, \mathbf{c}_1 \in \cc_1, \mathbf{c}_2 \in \cc_2\}.
\]
The code $\cc_1\oplus\cc_2$ is of parameters $\left[n_1+n_2,k_1+k_2,\min\{d_1,d_2\}\right]_q$. Let $K_1$ and $K_2$ be, respectively, parity-check matrices of $\cc_1$ and $\cc_2$. The block-diagonal matrix
\[
K_1 \oplus K_2 =
\begin{pmatrix}K_1& 0 \\ 0&K_2
\end{pmatrix}
\]
is a parity-check matrix of $\cc_1\oplus\cc_2$. We obtain a family of optimal $(r,\delta)$-LRCs of large lengths via the direct sum construction on a class of MDS codes and the linear codes designed in Theorem \ref{thm3}.

\begin{thm}\label{thm4}
Let $r$, $\delta$, $v > 1$, and $m := r+\delta-1 \leq q$ be positive integers. Let $\mathcal{H}$ be a ${v(\delta-1)\times (v(r+\delta-1)-1)}$ matrix defined by
\[
\mathcal{H}=H\oplus\cdots\oplus H\oplus\widehat{H},
\]
where $H$ and $\widehat{H}$ are the respective matrices in (\ref{m1}) and (\ref{m2}). Let $\cc$ be a linear code over $\Ff_q$ whose parity-check matrix is $\mathcal{H}$. Then $\cc$ is an optimal $(r,\delta)$-LRC with parameters $[vm-1,vr-1,\delta]_q$.
\end{thm}
\begin{IEEEproof}
Analogous to the proof of Theorem \ref{thm3}, we confirm that $\cc$ is an $(r,\delta)$-LRC. The direct sum construction implies that $\cc$ has parameters $[vm-1,vr-1,\delta]_q$, reaching the upper bound in (\ref{LRCbound}).
\end{IEEEproof}

\begin{example}
Let $r=2$, $\delta=3$, $q=5$, and $v=3$. We choose
\[
H=\begin{pmatrix}
1 & 1 & 1 & 1 \\
1 & 2 & 3 & 4
\end{pmatrix} \mbox{ and }
\widehat{H}=\begin{pmatrix}
1 & 1 & 1 & 1 & 0 & 0 & 0 \\
1 & 2 & 3 & 4 & 0 & 0 & 0 \\
0 & 0 & 0 & 1 & 1 & 1 & 1\\
0 & 0 & 0 & 4 & 3 & 2 & 1
\end{pmatrix}.
\]
Using $\mathcal{H}=H\oplus\widehat{H}$ as a parity-check matrix of $\mathcal{D}$ means that the parameters are $[11,5,3]_5$. The parameters of its punctured codes $\cc|_{\{1,2,3,4\}}$, $\cc|_{\{5,6,7,8\}}$, and $\cc|_{\{8,9,10,11\}}$ are all $[4,2,3]_5$. Thus, $\mathcal{D}$ is an optimal $(2,3)$-LRC.
\end{example}

To utilize Theorems \ref{thm2} and \ref{thm4}, we must construct a subcode of $\cc$, whose minimum distance is greater than $\delta$, as defined by Theorem \ref{thm4}. A common technique which preserves $(r,\delta)$-locality is to add well-chosen rows to the parity-check matrix $\mathcal{H}$ of $\cc$. Keeping the same notation as above, we write
\[
A= \left(\lambda_1f_{[\delta-1,\delta-2+\tau]} (\alpha_1),\, \ldots,\, \lambda_mf_{[\delta-1,\delta-2+\tau]}(\alpha_m)\right) \mbox{ and }
\widehat{A}=
\begin{pmatrix} A_1 & A_2
\end{pmatrix},
\]
with the two matrices over $\Ff_q$, namely, the $\tau \times m$ matrix $A_1$ and the $\tau\times (m-1)$ matrix $A_2$, given by
\begin{align*}
A_1 &:=\left(\lambda_1f_{[\delta-1,\delta-2+\tau]} (\alpha_1),\ldots, \lambda_mf_{[\delta-1,\delta-2+\tau]} (\alpha_m)\right) \mbox{ and } \\
A_2 &:=\left(\lambda_{m-1}f_{[\delta-1,\delta-2+\tau]} (\alpha_{m-1}),\,\ldots, \, \lambda_1f_{[\delta-1,\delta-2+\tau]}(\alpha_1)\right).
\end{align*}
Vertically joining $\mathcal{H}$ and $\begin{pmatrix}A & \cdots & A &\widehat{A} \end{pmatrix}$ yields a new $(v(\delta-1)+\tau)\times(vm-1)$ matrix
\begin{equation}\label{eq1}
\widetilde{\mathcal{H}}=\begin{pmatrix}
H & & &\\
  &\ddots & &\\
   &  & H & \\
  &  & & \widehat{H} \\
A & \cdots & A & \widehat{A}
\end{pmatrix}.
\end{equation}

\begin{thm}\label{thm5}
Let $r$, $\delta$, $\tau$, $v>1$ and $m :=r+\delta-1\leq q$ be positive integers such that $\tau+1<r$ and $\tau \leq \delta$. Assume that $\cc$ is a linear code over $\Ff_q$ with a parity-check matrix $\widetilde{\mathcal{H}}$ given by (\ref{eq1}). Then $\cc$ is an optimal $(r,\delta)$-LRC with parameters $[vm-1,vr-1-\tau,\delta+\tau]_q$.
\end{thm}
\begin{IEEEproof}
By how $\cc$ is defined and based on Theorem \ref{thm4}, $\cc$ has $(r,\delta)$-locality and length $vm-1$. Next, we determine its dimension. Choosing the last $(\delta-1+\tau)$ columns and the $(im+j)^{\rm th}$ column, for $i \in [v-1]$ and $j\in[\delta-1]$, we obtain an order $v(\delta-1)+\tau$ square matrix
\[
\widetilde{\mathcal{P}}=\begin{pmatrix}
P & & &\\
  &\ddots & &\\
   &  & P & \\
 B &\cdots  &B & \widehat{P}
\end{pmatrix},
\]
where
\begin{align*}
P &=\left(\lambda_1f_{[0,\delta-2]}(\alpha_1),\cdots,\lambda_{\delta-1}f_{[0,\delta-2]}(\alpha_{\delta-1})\right) \mbox{ and}\\
\widehat{P} &=\left(\lambda_{\delta-1+\tau}f_{[0,\delta-2+\tau]}(\alpha_{\delta-1+\tau}),\cdots,\lambda_{1}f_{[0,\delta-2+\tau]}(\alpha_{1})\right).
\end{align*}
Note that
\begin{align*}
\det(P) &=\lambda_1\cdots\lambda_{\delta-1}\prod_{1\leq s<t\leq \delta-1}(\alpha_s-\alpha_t)\neq 0 \mbox{ and}\\
\det(\widehat{P}) &= \lambda_1 \cdots \lambda_{\delta-1+\tau}\prod_{1\leq s<t\leq \delta-1+\tau}(\alpha_s-\alpha_t)\neq 0.
\end{align*}
Hence, $\det(\widetilde{\mathcal{P}})=\det(P)\cdots\det(P)\det(\widehat{P})\neq 0$, which implies $\rank(\widetilde{\mathcal{H}}) = v(\delta-1)+\tau$. Thus, the dimension of $\cc$ is $vr-1-\tau$. It follows from (\ref{LRCbound}) that the minimum distance
\[
d \leq  v(r+\delta-1)-1-(vr-1-\tau)-\left(\left\lceil\frac{vr-1-\tau}{r}\right\rceil-1\right)(\delta-1)+1 = \delta+\tau.
\]

By a method analogous to the one used in the proof of Theorem \ref{thm3}, any $(\delta-1+\tau)$ columns of $\widetilde{\mathcal{H}}$ are linearly independent over $\Ff_q$. Hence, the minimum distance of $\cc$ is equal to $\delta+\tau$. This completes the proof.
\end{IEEEproof}

We give a specific example to illustrate Theorem \ref{thm5}.

\begin{example}
Let $r=3$, $\delta=3$, $\tau=1$, $q=5$, and $v=3$. Let
\[
\widetilde{\mathcal{H}}=\begin{pmat}({....|.........})
1 & 1 & 1 & 1 & 1 & 0 & 0 & 0 & 0 & 0 & 0 & 0 & 0 & 0 \cr
0 & 1 & 2 & 3 & 4 & 0 & 0 & 0 & 0 & 0 & 0 & 0 & 0 & 0 \cr\-
0 & 0 & 0 & 0 & 0 & 1 & 1 & 1 & 1 & 1 & 0 & 0 & 0 & 0 \cr
0 & 0 & 0 & 0 & 0 & 0 & 1 & 2 & 3 & 4 & 0 & 0 & 0 & 0 \cr
0 & 0 & 0 & 0 & 0 & 0 & 0 & 0 & 0 & 1 & 1 & 1 & 1 & 1 \cr
0 & 0 & 0 & 0 & 0 & 0 & 0 & 0 & 0 & 4 & 3 & 2 & 1 & 0 \cr\-
0 & 1 & 2^2 & 3^2 & 4^2 & 0 & 1 & 2^2 & 3^2 & 4^2 & 3^2 & 2^2 & 1^2 & 0 \cr
\end{pmat}
\]
be a parity-check matrix of $\cc$. Here, we have
\[
H =\begin{pmatrix}
1 & 1 & 1 & 1 & 1\\
0 & 1 & 2 & 3 & 4
\end{pmatrix}, \quad
\widehat{H}=\begin{pmatrix}
1 & 1 & 1 & 1 & 1 &0 & 0 & 0 &0\\
0 & 1 & 2 & 3 & 4 & 0 & 0 & 0 & 0 \\
0 & 0 & 0 & 0 & 1 & 1 & 1 & 1 & 1\\
0 & 0 & 0 & 0 & 4 & 3 & 2 & 1 & 0
\end{pmatrix},
\]
\[
A =\begin{pmatrix}
0 & 1 & 2^2 & 3^2 & 4^2
\end{pmatrix}, \quad
\widehat{A} = \begin{pmatrix}
0 & 1 & 2^2 & 3^2 & 4^2 & 3^2 & 2^2 & 1^2 & 0
\end{pmatrix}.
\]
The parameters of $\cc$ are $[14,7,4]_5$. The punctured codes $\cc|_{\{1,2,3,4,5\}}$, $\cc|_{\{6,7,8,9,10\}}$ and $\cc|_{\{10,11,12,13,14\}}$ are all $[5,3,3]_5$. Thus, $\cc$ is an optimal $(3,3)$-LRC.
\end{example}

Theorem \ref{thm4} gives us a $[v  (r+\delta-1)-1,vr-1,\delta]_q$-optimal $(r,\delta)$-LRC $\cc_1$. The subcode of $\cc_1$ identified by Theorem \ref{thm5} is also an optimal $(r,\delta)$-LRC. Thus, Theorem \ref{thm2} yields the following family of optimal $(r,\delta)$-LRCs.

\begin{thm}\label{thm6}
Let $r$, $\delta$, $\tau$, and $v$ be positive integers such that
\[
v>1, \quad r+ \delta-1 \leq q, \quad
\tau + 1 < r \mbox{, and }\tau \leq \delta.
\]
Let $\cc_1= \cdots = \cc_{N-1}$ be optimal $(r,\delta)$-LRCs with parameters $[v(r+\delta-1)-1,vr-1,\delta]_q$ constructed by Theorem \ref{thm4}. Let $\cc_N$ be the subcode of $\cc_1$, with parameters $[v(r+\delta-1)-1,vr-1-\tau,\delta+\tau]_q$, designed by Theorem \ref{thm5}.
Let $\cc:=(\cc_1,\cdots,\cc_N)\cdot A$ be the matrix-product code constructed in (\ref{MPC}), with $A$ being an NSC matrix of order $N\leq q$. If $N+\tau <r$, then $\cc$ is an optimal $(r,\delta)$-LRC with parameters $[Nv(r+\delta-1)-N, Nvr-N-\tau, \delta+\tau]_q$.
\end{thm}
\begin{IEEEproof}
The assertion follows directly from Theorem \ref{thm2}.
\end{IEEEproof}

\begin{example}
Let $q=7$, $r=4$, $\delta=3$, $N=v=2$, and $\tau=1$. Let $\cc_1$ and $\cc_2$ be $7$-ary linear codes with respective parity-check matrices
\[
\mathcal{H}=
\begin{pmatrix}
1 & 1 & 1 & 1 & 1 & 1 & 0 & 0 & 0 & 0 & 0 \\
1 & 2 & 3 & 4 & 5 & 6 & 0 & 0 & 0 & 0 & 0 \\
0 & 0 & 0 & 0 & 0 & 1 & 1 & 1 & 1 & 1 & 1 \\
0 & 0 & 0 & 0 & 0 & 6 & 5 & 4 & 3 & 2 & 1
\end{pmatrix}
\mbox{ and }
\widetilde{\mathcal{H}}=
\begin{pmatrix}
1 & 1 & 1 & 1 & 1 & 1 & 0 & 0 & 0 & 0 & 0 \\
1 & 2 & 3 & 4 & 5 & 6 & 0 & 0 & 0 & 0 & 0 \\
0 & 0 & 0 & 0 & 0 & 1 & 1 & 1 & 1 & 1 & 1 \\
0 & 0 & 0 & 0 & 0 & 6 & 5 & 4 & 3 & 2 & 1 \\
1 & 2^2 & 3^2 & 4^2 & 5^2 & 6^2 & 5^2 & 4^2 & 3^2 & 2^2 & 1
\end{pmatrix}.
\]
The codes $\cc_1$ and $\cc_2$ are optimal $(4,3)$-LRCs with respective parameters $[11,7,3]_7$ and $[11,6,4]_7$ and  generator matrices
\[
G_1=
\begin{pmatrix}
1&0&0&0&2&4&0&0&0&1&2 \\
0&1&0&0&3&3&0&0&0&6&5 \\
0&0&1&0&4&2&0&0&0&4&1 \\
0&0&0&1&5&1&0&0&0&2&4 \\
0&0&0&0&0&0&1&0&0&3&3 \\
0&0&0&0&0&0&0&1&0&4&2 \\
0&0&0&0&0&0&0&0&1&5&1
\end{pmatrix}
\mbox{ and }
G_2=
\begin{pmatrix}
1&0&0&0&2&4&0&0&1&6&3\\
0&1&0&0&3&3&0&0&4&5&2\\
0&0&1&0&4&2&0&0&6&6&0\\
0&0&0&1&5&1&0&0&0&2&4\\
0&0&0&0&0&0&1&0&1&1&4\\
0&0&0&0&0&0&0&1&4&3&6
\end{pmatrix}.
\]
Let $A=
\begin{pmatrix}
1 & 1\\
1 & 2
\end{pmatrix}$. The code $\cc=(\cc_1,\cc_2)\cdot A$ has a generator matrix $\begin{pmatrix}G_1 & G_1\\ G_2 & 2 \, G_2\end{pmatrix}$ and parameters $[22,13,4]_7$. Each of the punctured codes $\cc|_{\{1,2,3,4,5,6\}}$, $\cc|_{\{6,7,8,9,10,11\}}$, $\cc|_{\{12,13,14,15,16,17\}}$ and $\cc|_{\{17,18,19,20,21,22\}}$ has parameters $[6,4,3]_7$. Thus, $\cc$ is an optimal $(4,3)$-LRC.
\end{example}

\begin{figure}[ht!]
\centering
\caption{Comparison of the parameters $[n,k,d]_q$ of known and new LRCs listed in Table~\ref{tabl1} for the stated $r$, $\delta$, $q$, and $n$. One can clearly see the new ones in relation to those already known.}
\begin{tabular}{cc}
\includegraphics[width=0.47\linewidth]{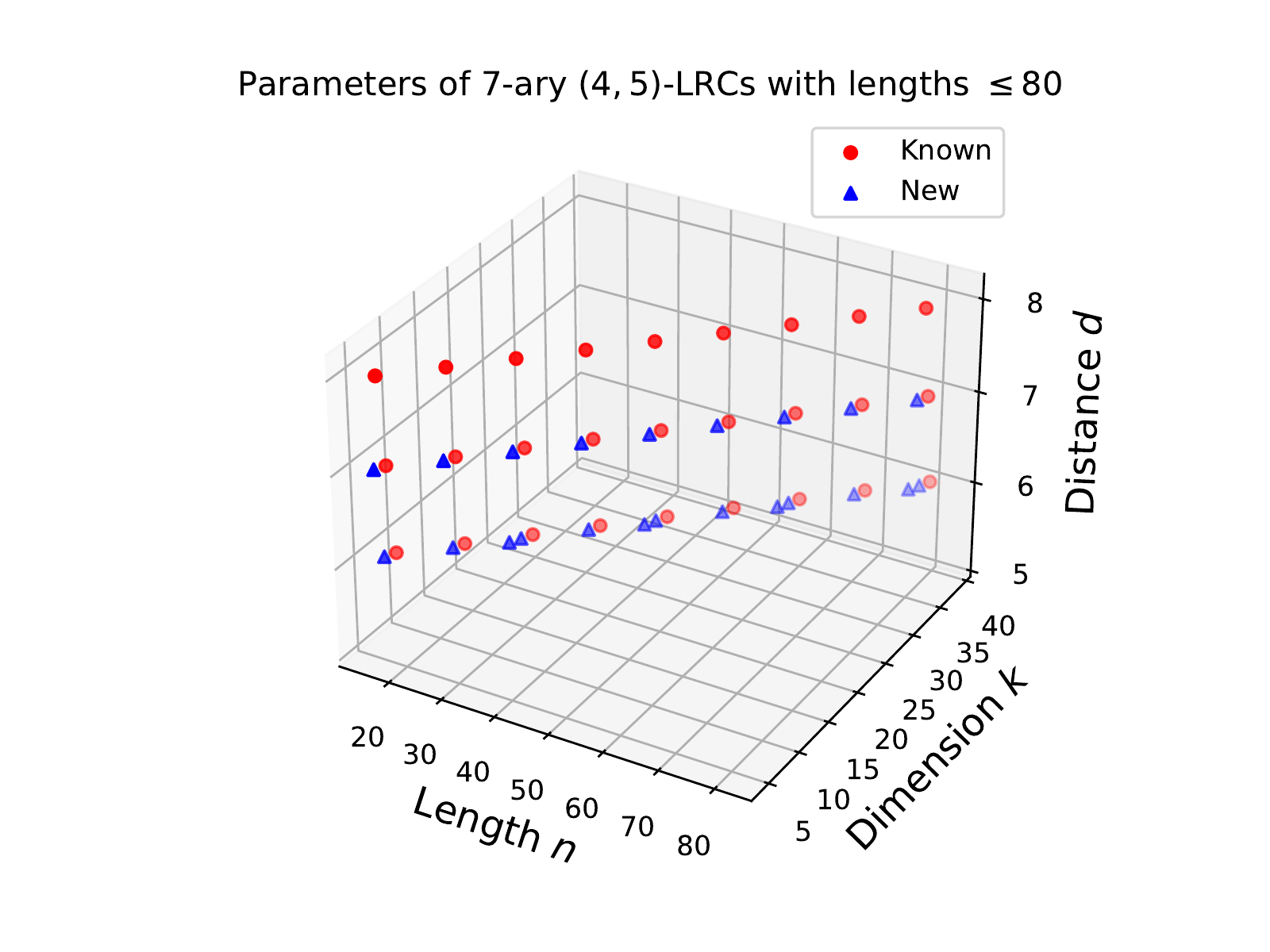} &
\includegraphics[width=0.47\linewidth]{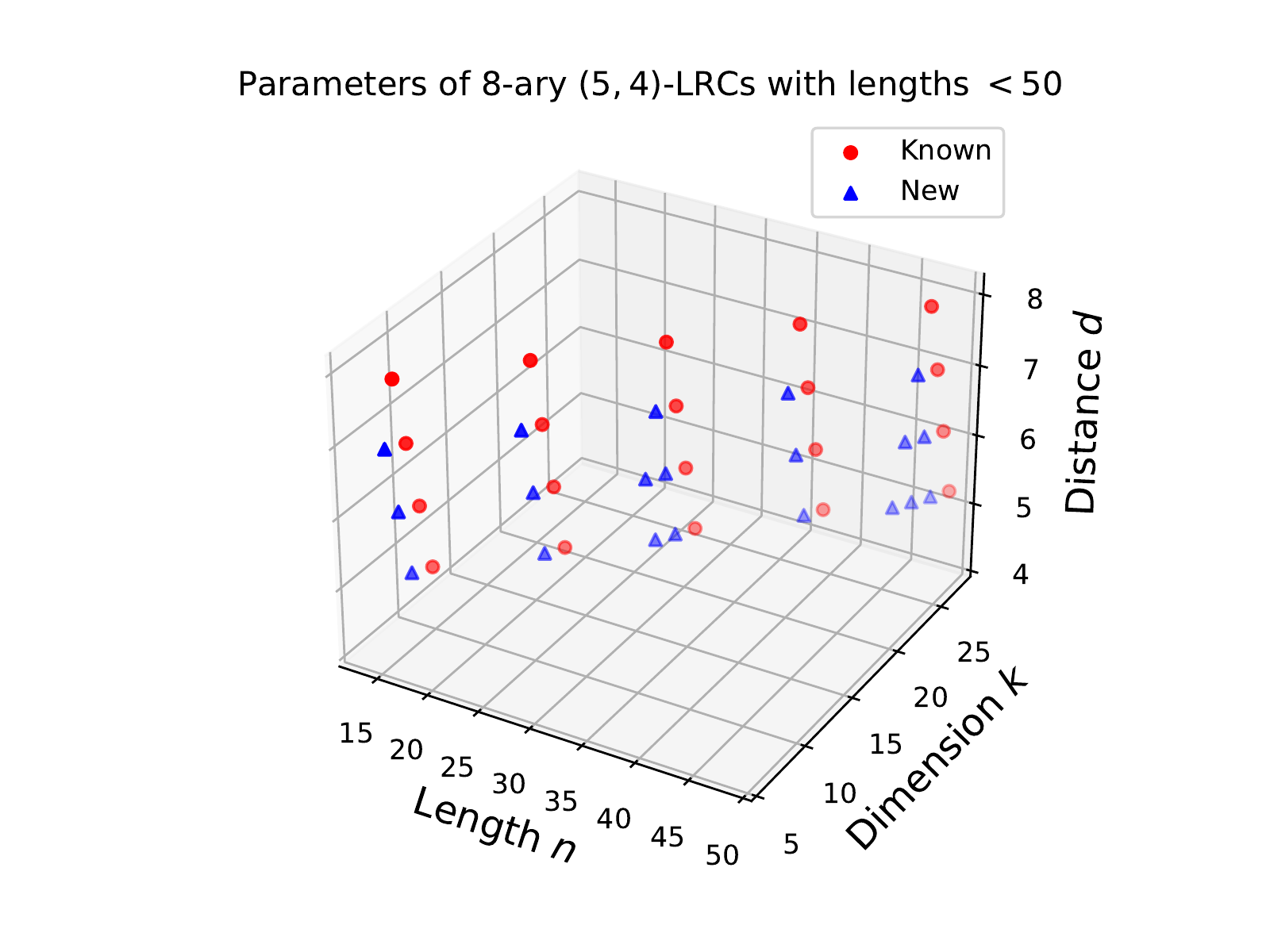} \\
(a) When $(q,r,\delta,n)=(7,4,5,\leq 80)$. & (b) When $(q,r,\delta,n)=(8,5,4,<50)$.
\end{tabular}
\label{fig:illust}
\end{figure}

\begin{remark}\label{cp2}
An infinite family of optimal $(r,\delta)$-LRCs with unbounded lengths and parameters $[n=w(r+\delta-1),wr-s,\delta+s]_q$, where $\delta+s\leq r+\delta-1 \leq q$ and $s\leq \delta$, can be found in \cite{Cai,ChenBC}. Utilizing cyclic codes, Fang and Fu \cite{Fang} presented three families of optimal $(r,\delta)$-LRCs over $\Ff_q$ with unbounded lengths and minimum distance $d$, with $\delta\leq d\leq 2\delta$. The lengths of these known optimal $(r,\delta)$-LRCs are divisible by $(r+\delta-1)$. Note that the optimal $(r,\delta)$-LRCs generated by Theorem \ref{thm6} have lengths not divisible by $(r+\delta-1)$ and minimum distance $d$ with $\delta\leq d\leq 2\delta$. Our constructions of optimal $(r,\delta)$-LRCs are, therefore, new.

For reference, the parameters of known optimal $(r,\delta)$-LRCs with unbounded lengths and the new ones produced by Theorem \ref{thm6} are listed in Table \ref{tabl1}. Figure~\ref{fig:illust} provides two sets of scattered plots that compare the parameters $[n,k,d]_q$ of previously known LRCs in Entries 1 to 4 of Table~\ref{tabl1} with the new ones in Entry 5 for the specified $r$, $\delta$, $q$, and range of lengths.
\end{remark}

\begin{table}[htb!]
\renewcommand{\arraystretch}{1.2}
\caption{Parameters of optimal $(r,\delta)$-LRCs with unbounded lengths over $\Ff_q$, with $ m :=r+\delta-1$ and $v>1$.}
\label{tabl1}
\renewcommand{\arraystretch}{1.2}
\centering
\begin{tabular}{ccccll}
\toprule
No. & Length $n$ & Dimension $k$ & Distance $d$ & Constraints & References \\ \midrule
$1$ & $v m $& $v r-s$ & $\delta+s$ & $\delta+s\leq  m  \leq q, \quad s\leq \delta, \quad  m  \mbox{ divides } n$ & \cite{Cai,ChenBC}  \\% \hline

$2$ &  $v m $& $v r-1$ & $\delta+1$ & $\gcd(n,q)=1, \quad r\geq 2, \quad  m  \mbox{ divides } \gcd(n,q-1)$ & \cite{Fang}  \\

$3$ & $v m $& $v r-2$ &$\delta+2$ & $\gcd(n,q) =1, \quad r\geq 3, \quad \gcd\left(\frac{n}{ m }, m \right) \mbox{ divides } \delta$, & \cite{Fang}  \\
& & & & $ m  \mbox{ divides } \gcd(n,q-1)$ & \\
%\hline

$4$ & $v m $& $v r-\delta$ &$2\delta$ & $\gcd(n,q) =1, \quad
3\leq \delta+1\leq r, \quad \gcd\left(\frac{n}{ m }, m \right)=1$, & \cite{Fang} \\
&&&& $ m  \mbox{ divides } \gcd(n,q-1)$ & \\ %\hline

$5$ & $N\mu  m -N$ & $N\mu r-N-\tau$ & $\delta+\tau$ & $ N \geq 1, \quad \mu>1, \quad N+\tau <r, \quad \tau \leq \delta, $ & Thms. \ref{thm5} and \ref{thm6}  \\
&&&& $\delta+\tau\leq  m  \leq q$ & \\ %\hline
\bottomrule
\end{tabular}
\end{table}

\section{Concluding remarks}

We have investigated the locality of the matrix-product code $\cc=(\cc_1,\ldots,\cc_M) \cdot A$ and demonstrated that the matrix-product code $\cc$ preserves the $(r,\delta)$-locality of $\cc_1$, if $\cc_M\subseteq \cdots \subseteq\cc_1$. A major advantage of our approach lies in its simplicity. An $(r,\delta)$-LRC of length $n$, with $n$ large, can be built from a smaller $(r,\delta)$-LRC and its nested codes. Most classical linear codes of large length $n$ have complicated structures. This complexity in construction carries over to most prior constructions of optimal $(r,\delta)$-LRCs, since the latter are constructed based on the parity-check matrices of the underlying classical codes.

Our approach, via matrix-product codes, establishes the properties of relevant linear codes of large lengths based on the properties of linear codes of small lengths. Our first two families of optimal $(r,\delta)$-LRCs with superlinear length are built from GRS codes and cyclic or constacyclic MDS codes. By designing optimal $(r,\delta)$-LRCs with unbounded length $n$, provided that $n$ is not a multiple of $(r+\delta-1)$, we have come up with the third family of optimal $(r,\delta)$-LRCs. The codes in this family have flexible parameters due to the versatility of the matrix-product operation.

\end{document}